\documentclass[10pt,conference]{IEEEtran}
\IEEEoverridecommandlockouts
\usepackage{filecontents,lipsum}
\usepackage[noadjust]{cite}
\usepackage{amsmath,amssymb,amsfonts}
\usepackage{algorithmic}
\usepackage{graphicx}
\usepackage{textcomp}
\usepackage{xcolor}

\def\BibTeX{{\rm B\kern-.05em{\sc i\kern-.025em b}\kern-.08em T\kern-.1667em\lower.7ex\hbox{E}\kern-.125emX}}
    
\begin{document}

\title{What do pre-trained code models know about code?}

\author{
\IEEEauthorblockN{Anjan Karmakar}
\IEEEauthorblockA{\textit{Free University of Bozen-Bolzano} \\
Bolzano, Italy \\
akarmakar@unibz.it}
\and
\IEEEauthorblockN{Romain Robbes}
\IEEEauthorblockA{\textit{Free University of Bozen-Bolzano} \\
Bolzano, Italy \\
rrobbes@unibz.it}
}

\maketitle

\begin{abstract}
Pre-trained models of code built on the transformer architecture have performed well on software engineering (SE) tasks such as predictive code generation, code summarization, among others. However, whether the vector representations from these pre-trained models comprehensively encode characteristics of source code well enough to be applicable to a broad spectrum of downstream tasks remains an open question. 

One way to investigate this is with diagnostic tasks called \textit{probes}. In this paper, we construct four probing tasks (probing for surface-level, syntactic, structural, and semantic information) for pre-trained code models. We show how probes can be used to identify whether models are deficient in (understanding) certain code properties, characterize different model layers, and get insight into the model sample-efficiency. 

We probe four models that vary in their expected knowledge of code properties: \texttt{BERT} (pre-trained on English), \texttt{CodeBERT} and \texttt{CodeBERTa} (pre-trained on source code, and natural language documentation), and \texttt{GraphCodeBERT} (pre-trained on source code with dataflow). While \texttt{GraphCodeBERT} performs more consistently overall, we find that \texttt{BERT} performs surprisingly well on some code tasks, which calls for further investigation. 

\end{abstract}


\section{Introduction}

The outstanding success of transformer-based \cite{vaswani2017attention} pre-trained models in NLP such as \texttt{BERT} \cite{devlin2019bert}, has inspired the creation of a number of similar pre-trained models for source code
\cite{DBLP:journals/corr/abs-1910-03771, lachaux2020unsupervised, svyatkovskiy2020intellicode, clement2020pymt5, tufano2020generating, ahmad2020transformerbased}. 
These pre-trained models are first trained on a large corpus of code in a self-supervised manner and then fine-tuned on downstream tasks.

The progress made with pre-trained source code models is genuinely encouraging with applications in software security, software maintenance, software development and deployment. And although the pre-trained vector embeddings from the transformer models have worked well on many tasks, it remains unclear what exactly these models learn about code---specifically what aspects of code structure, syntax, and semantics are known to it. Thus, our work is motivated by the need to assess the properties of code that are learned by pre-trained embeddings, in order to build accurate, robust, and generalizable models for code, beyond single-task models. 

An emerging field of research addresses this objective by means of \textit{probes}. Probes are diagnostic tasks, in which a simple classifier is trained to predict specific properties of its input, based on the \emph{frozen} vector embeddings of a pre-trained model. The degree of success in the probing tasks indicates whether the information probed for is present in the pre-trained embeddings. Probing has been extensively used for natural language models, and has already begun to pick up steam with numerous probing tasks \cite{alain2018understanding, shi-etal-2016-string, adi2017finegrained, belinkov-etal-2017-neural,  DBLP:journals/corr/abs-1808-08949, DBLP:journals/corr/abs-1905-06316, conneau-etal-2018-cram, tenney2019learn} investigating a diverse array of natural language properties. 

In this work, we adapt the probing paradigm to pre-trained source code models. We assess the hidden state embeddings of multiple models, and determine their ability to capture elemental characteristics related to code, that may be suitable for use in several downstream SE tasks. We evaluate \texttt{CodeBERT} \cite{feng2020codebert}, \texttt{CodeBERTa} \cite{DBLP:journals/corr/abs-1910-03771}, and \texttt{GraphCodeBERT} \cite{guo2021graphcodebert}, with \texttt{BERT} \cite{devlin2019bert} as our transformer baseline. As an initial study, we have chosen to work with \texttt{BERT} and its code-trained descendants, as it provides a ground for comparison among natural language models, models trained jointly on natural language and code, models trained just on code, and models trained on code with additional structural information. 


We construct four initial probing tasks for this purpose: {AST Node Tagging}, {Cyclomatic Complexity Prediction}, {Code Length Prediction}, and {Invalid Type Detection}. These four tasks are meant to assess whether pre-trained models are able to capture different aspects of code, specifically the syntactic, structural, surface-level and semantic information respectively. The tasks were chosen to cover the most commonly identifiable abstractions of code, although more tasks are needed to thoroughly probe for each type of code abstraction.



This paper makes the following contributions:
\begin{itemize}
    \item An introduction to probing for pre-trained code models. We introduce four probing tasks each probing a particular characteristic of code, and release the corresponding task datasets publicly.
    \item A preliminary empirical study, based on probing tasks and pre-trained code models, that highlights the potential of probes as a pseudo-benchmark for pre-trained models.
    \item A discussion on the efficacy of pre-trained models. We show to what extent different code properties are encoded in pre-trained models. 
\end{itemize} 


Overall, our probes suggest that the models do encode the syntactic and semantic properties, to varying degrees. While we find that models that have more knowledge of source code tend to perform better at the more code-specific probing tasks, yet, the difference in performance between the baseline and the source code models are smaller than expected. This calls for further study of the phenomenon, and for increased effort in designing pre-training procedures that better capture diverse source code characteristics. 




\section{Background}

A probe fundamentally consists of a probing task and a probing classifier. A probing task is an auxiliary diagnostic task that is constructed to determine whether a specific property is encoded in the pre-trained model weights. Probing is useful when assessing the raw predictive power of pre-trained weights without any sort of fine-tuning with (downstream) task data. Probing tasks are often simple in nature compared to downstream tasks to minimize interpretability problems. 




A probing classifier, on the other hand, is used to train on the probing task where the input vectors of the training samples are extracted from the \emph{frozen} hidden layers of the pre-trained model. Importantly, the probing classifier, which is usually a linear classifier, is \emph{simple} with no hidden layers of its own. If a simple probing classifier can predict a given attribute from the pre-trained embeddings, the original model most likely encodes it in its hidden layers. Usually, the raw accuracies from a probe are not the focus of the study; rather, the probe is used to assesses whether a model encodes a characteristic \emph{better} than another, or compares several model layers.


\textit{Related work.} Studies in NLP research have shown how several pre-trained natural language models encode different linguistic properties such as sentence length, and verb tense, among other properties \cite{conneau-etal-2018-cram}. Studies such as \cite{jawahar-etal-2019-bert} show that \texttt{BERT} encodes phrase-level information in the lower layers, and a hierarchy of linguistic information in the intermediate layers, with surface features at the bottom, syntactic features in the middle and semantic features at the top of a vector space. 
Other studies focus on word morphology \cite{belinkov-etal-2017-neural}, or syntax \cite{shi-etal-2016-string}, to name a few.  Studies of the \texttt{BERT} models alone have spawned a subfield known as \textit{\texttt{BERT}ology} with over 150 studies surveyed \cite{rogers2020primer}. While probing is well established in NLP, it is almost absent for source code models. The only example we are aware of uses a single coarse task (programming language identification)---and is not the focus of the paper \cite{feng2020codebert}.




\section{Probing Source Code}

\textbf{Probing Tasks.}
In order to determine whether the pre-trained vector embeddings of  source code transformer models reflect code understanding in terms of syntactic, semantic, structural, and surface-level characteristics, we have constructed four diverse probing tasks. 

\textit{AST Node Tagging (AST)}
As Abstract Syntax Trees (ASTs) are the basis of many structured source code representations \cite{alon2018code2vec, alon2019code2seq, DBLP:journals/corr/abs-1711-00740, brockschmidt2019generative, wei2020lambdanet, ijcai2017-423, mou2015convolutional, DBLP:conf/iclr/HellendoornSSMB20}, they emerge as a rational choice to evaluate pre-trained source code models on syntactic understanding. In order for a pre-trained code model to be good at code tasks such as code completion, it must necessarily learn and interpret the syntactic structure of a sequence of code tokens and predict a syntactically valid next token. Thus, identifying AST node tags often is a hidden prerequisite to solving a given code task---making it a suitable contender for probing any pre-trained source code model. We probe the pre-trained models with this task to determine to what extent \textbf{syntactic information} is encoded in the model layers. 

\textit{Cyclomatic Complexity (CPX)}
To probe whether some sort of code \textbf{structure information} is encoded in the hidden layers of a pre-trained model, we construct the cyclomatic complexity task. Since the complexity is an inherent characteristic of any code snippet, the models should be able to predict it without explicit fine-tuning. Furthermore, since the complexity of a code snippet depends on the number of linearly independent paths through the code snippet, predicting it based simply on the sequence of tokens might be a challenge.

\textit{Code Length Prediction (LEN)}
We conjecture that the length of a code snippet, especially when it is fed into the model as a sequence of code tokens, should be easy to predict for the transformer models. To determine whether the code transformers encode such elementary \textbf{surface information}, we probe the models with the code length prediction task. 

\textit{Invalid Type Detection (TYP)}
To understand to what extent pre-trained code models are aware of code semantics and are able to distinguish between semantically valid snippets of code from invalid ones---keeping both syntactically legitimate, we construct an invalid {type} detection task. For this probing task, the negative samples consist of code snippets where some \textit{primitive data types} are misspelled intentionally. The code transformers are probed to determine if they are able to classify code snippets based on invalid \textbf{semantic information}. 

\textit{Probing data \& labels.} We gather our data from a subset of the 50K-C dataset of compilable Java projects \cite{8595165_50KC}. We construct ASTs of method-level code snippets from several of the largest projects and collect a diverse range of AST node tags as labels: totaling 20 classes of node type labels. We use the metrix++ tool to obtain the complexity labels (with complexities ranging from 0 to 9). We tokenize our training samples with ANTLR to gather the length labels in five class-bins (0-50, 50-100, etc.). The labels for invalid types are obtained by interchanging consecutive characters at random indexes to resemble misspelled types. Our probing datasets have 10k samples for each task and are class-balanced.






\textbf{Pre-trained Models.}
We probe four state-of-the-art models: \texttt{BERT}, \texttt{CodeBERT}, \texttt{CodeBERTa}, and \texttt{GraphCodeBERT}. All the models are built upon the multi-layer bidirectional transformer introduced by Vaswani et al. \cite{vaswani2017attention}. Other than \texttt{BERT}, which is our baseline, pre-trained on a large corpus of English data, all the other transformer models are pre-trained on the CodeSearchNet dataset extracted from GitHub \cite{husain_codesearchnet_2019}, which includes 6.4 million methods across six programming languages. While \texttt{CodeBERT} and \texttt{CodeBERTa} gain knowledge of code by training on source code and natural language documentation, \texttt{GraphCodeBERT} goes further, encoding even data-flow information extracted from the ASTs.

\textbf{Probing Classifier.} We train a simple linear classifier that takes the input feature vectors from the hidden layers of a pre-trained code transformer. This is done to determine which of the code properties are linearly correlated with pre-trained model embeddings. Since a linear classifier has a basic model architecture with no hidden units, therefore it must heavily rely on the pre-trained embeddings to do well in the tasks.

\section{Early Results and Discussion} 

The results and observations from the probe analyses are discussed below. It is important to note that our interest is more on the difference between the models rather than the general accuracy of the probes. 


\begin{table}[htbp]
  \caption{Probing Task Accuracies}
  \label{tab:commands}
  \begin{tabular}{lcccc}
   \hline & \\[-1.5ex]
                          & LEN               & AST                & CPX                 & TYP               \\
    \texttt{Model}        & \textit{surface}  & \textit{syntactic} & \textit{structural} & \textit{semantic} \\
    \hline & \\[-1.5ex]
    \texttt{Naive}        & 20.00             & 05.00              & 10.00               & 50.00             \\
    \\[-1.2ex]
    \texttt{BERT}         & \textbf{76.05}    & \underline{89.90}  & \underline{42.65}   & 86.85             \\
    \texttt{CodeBERT}     & 68.15             & 89.45              & 41.40               & \underline{93.85} \\
    \texttt{CodeBERTa}    & 70.35             & \textbf{92.55}     & 40.80               & 90.10             \\
    \texttt{GraphCodeBERT}& \underline{71.10} & 85.50              & \textbf{46.70}      & \textbf{97.20}    \\
    [0.8ex]
    \hline 
  \end{tabular}
\end{table}


\subsection{Model Analysis}
\textit{\textbf{Surface-level information.}} \texttt{BERT} performs the best with 76.05\% accuracy for code length prediction. Considering the amount of training data \texttt{BERT} has seen and given the task of predicting the number of tokens from the input sequence, it is not unexpected that \texttt{BERT} does well on this task. 

\textit{\textbf{Syntactic information.}} For AST node tagging, we find that with enough training samples, all of the pre-trained models are able to achieve classification accuracies beyond 85\%, reaching up to 92.55\% for \texttt{CodeBERTa} \cite{DBLP:journals/corr/abs-1910-03771}. This shows that syntax-related information requisite for the node tagging task is encoded in the model hidden layers.

\textit{\textbf{Structural information.}} \texttt{GraphCodeBERT}, having been trained with the most structured information, performs the best with 46.70\% accuracy for cyclomatic complexity prediction. The standard \texttt{BERT} model also performs well with 42.65\% accuracy. This raises the question whether \texttt{BERT} encodes structural information of code, or alternatively, whether the pre-trained code models fail to encode it effectively. 

\textit{\textbf{Semantic information.}} \texttt{GraphCodeBERT} has the highest accuracy (97.20\%) for invalid type prediction. Although all of the pre-trained code models exhibit prediction accuracies beyond 90\%, \texttt{BERT} is not far behind in terms of accuracy. Note that since this task is a simple binary classification task, a higher naive baseline accuracy is expected. 

\textit{\textbf{Overall observations.}} In all cases, the hidden-layer vector embeddings of the pre-trained models seem to encode information that correlates with code characteristics to varying degrees. The structure-based code task (CPX) is the most challenging: implying that the pre-trained embeddings may not have direct linear correlations with the probed code properties, as linear classifiers are not able to extract it effectively.



While \texttt{GraphCodeBERT} has the most consistent performance overall, 
it is surprising that \texttt{BERT}, pre-trained just on English text, exhibits a similar competitive performance against the other pre-trained code models---which \emph{should} have more knowledge of code. \texttt{BERT} does considerably well on most tasks, even for the tasks which are more code-dependent, which calls for an extended investigation on this topic.   


\subsection{Layer analysis}
\begin{figure}[h]
  \centering
  \includegraphics[width=0.95\linewidth]{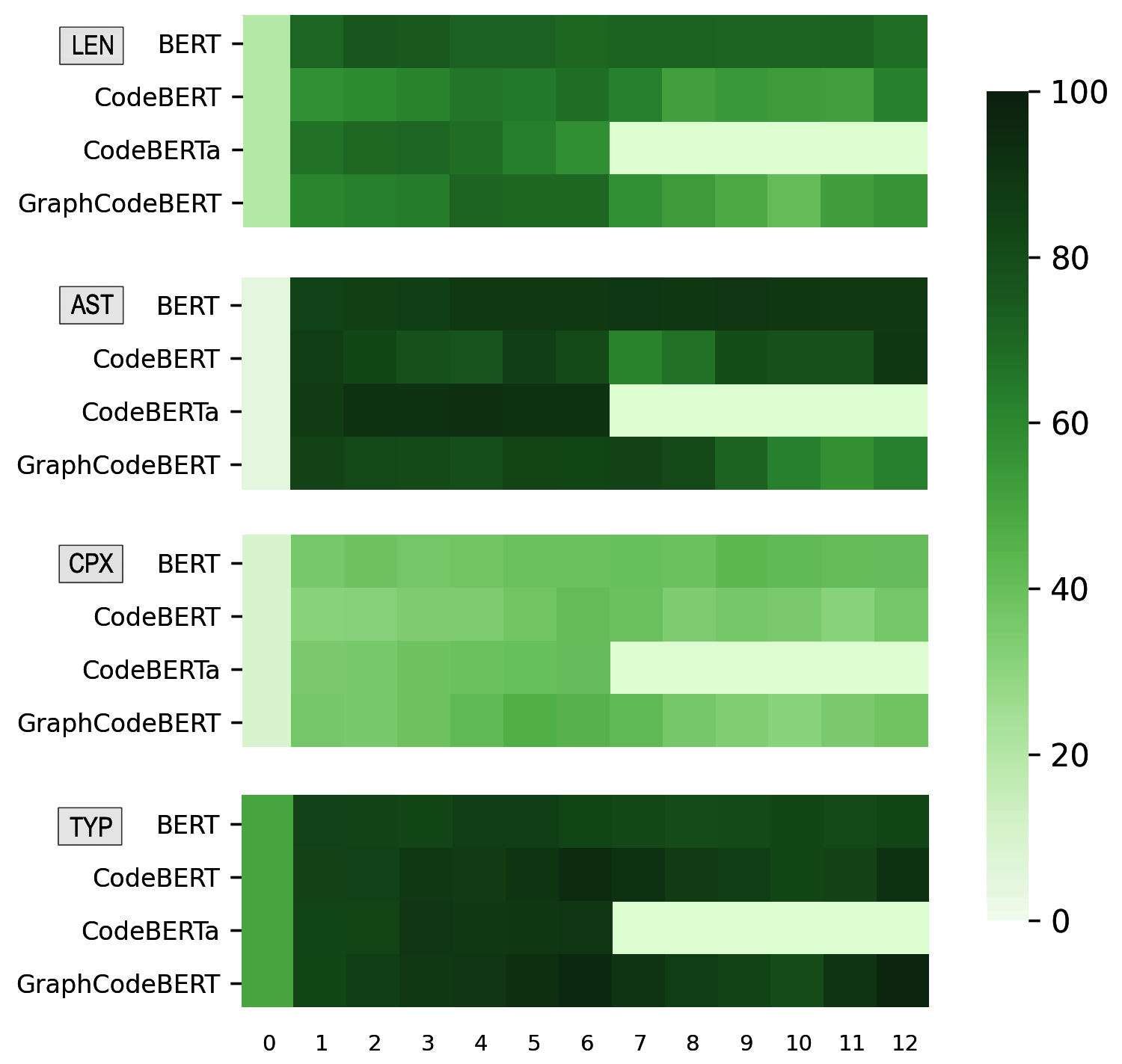}
  \caption{Pre-trained model accuracies by layers.} 
  \label{fig:layer_analysis}
\end{figure}

Alain \& Bengio \cite{alain2018understanding} compare \textit{probes} to 
thermometers used to measure the temperature (accuracies) simultaneously at several different locations (layers). Figure \ref{fig:layer_analysis} shows the accuracies of the pre-trained source code models across all its layers. 

All of the models have 12 layers, except \texttt{CodeBERTa} which has only 6. The accuracies are based on the pre-trained vector embeddings extracted from each of these hidden layers ranging from 1-12. The encoding layer of each model is represented by layer 0 which displays naive baseline accuracy. 



The pre-trained code models appear to have heterogeneous performance across the layers. We observe no single layer consistently performs optimally for all tasks, which is expected, indicating that the abstraction of the different learned code properties are spread across multiple layers.

BERT shows similar behavior for English \cite{jawahar-etal-2019-bert}, but it does not show this localization of performance for code---exhibiting fairly homogeneous performance across the layers. 






\subsection{Sample Size Analysis}

Pre-trained models when fine-tuned on a downstream task, need only a fraction of the data a model trained from scratch needs to do well. Hence, when evaluating pre-trained models with probes it is also essential to study the effects of training data volume. Thus we evaluate how the probes perform when data is scarce - with 10\% and 1\% of the dataset size. 

We limit our training to a maximum of 10,000 samples with the intuition that the general underlying syntactic, semantic, structural rules can be learned in a sample-efficient way and does not require memorization.


\begin{figure}[h]
  \centering
  \hbox{\hspace{-0.5ex}\includegraphics[width=1\linewidth]{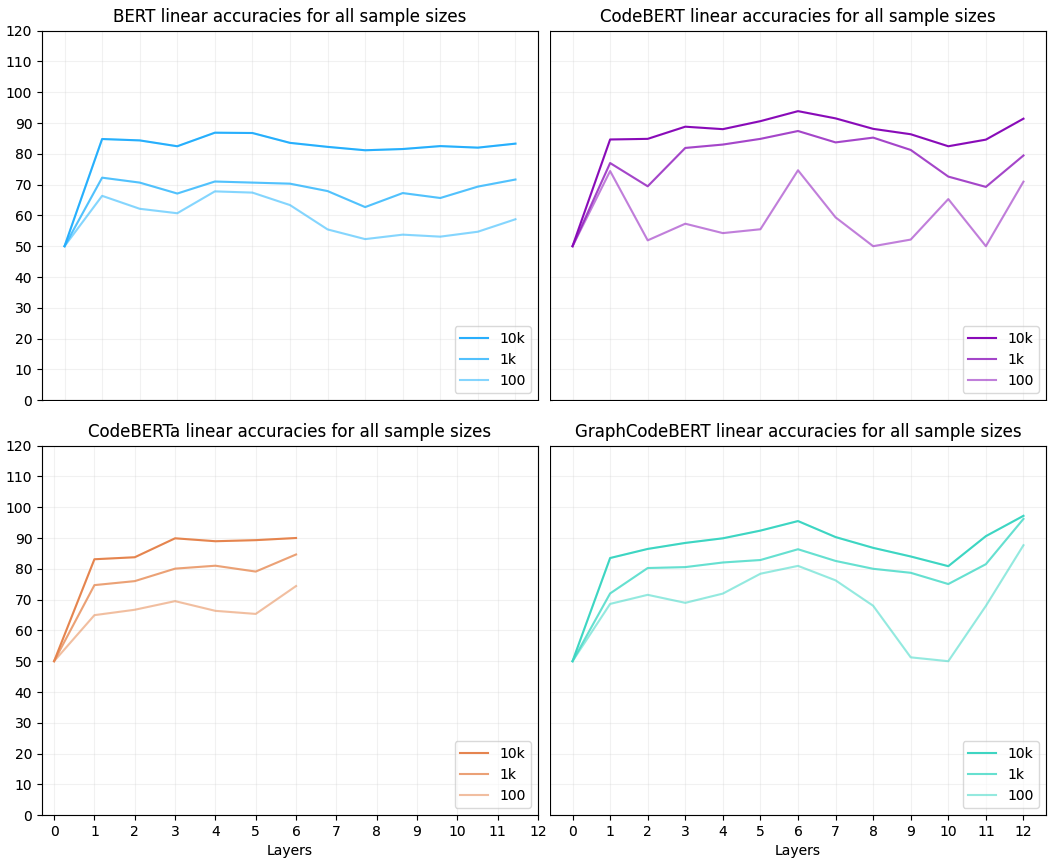}}
  \caption{Model accuracies by sample sizes for \textit{Invalid Type Prediction}}
  \label{fig:models_acc_samples_legend}
\end{figure}

As expected, the overall model accuracies increase with the increase in the number of samples from 100 to 1000 to 10,000 samples. However, it is interesting to note that the irregularity (uneven trend) in accuracies from layer to layer is flattened as more samples are provided (Fig. \ref{fig:models_acc_samples_legend}). This applies to all tasks. 


An intriguing observation is that \texttt{GraphCodeBERT}'s performance with just 100 samples exceeds \texttt{BERT}'s performance with 10,000 samples for the TYP task. 
We attribute this to \texttt{GraphCodeBERT} being much more sample-efficient for this task, as it can extract a lot of the signal from fewer samples. For the other tasks, such as AST and CPX where pre-trained code models improved upon the baseline, no such patterns implying sample-efficiency were present. 








\subsection{Discussion}

\textbf{\textit{Code properties.}} Our findings suggest that while certain code characteristics can be extracted from pre-trained code transformers with linear classifiers, implying that they are firmly encoded in the hidden layers, others, such as cyclomatic complexity, cannot be extracted as effectively. We plan to investigate whether these characteristics can be extracted with non-linear classifers such as an MLP with its own hidden layers \cite{conneau-etal-2018-cram}. In general, the results also point to the idea that more effective pre-training procedures that can make these characteristics more accessible should be explored.


\textbf{\textit{Model performance.}} For all of the tasks \texttt{BERT} has shown competitive performance against the other \texttt{BERT}-style code transformers. It shows understanding of surface information better than the others, while being the next best model for syntax- and structure-based evaluations. \texttt{CodeBERTa} and \texttt{CodeBERT} show promising results for syntax and semantic understanding respectively, with an improvement on the baseline \texttt{BERT} model. \texttt{GraphCodeBERT} struggles to reach accuracies beyond 85\% on AST Node Tagging task---we hypothesize it is missing out on the data-flow context while making predictions based on a single code token. Besides that, it is the most consistent model with up to 10-12\% improvements in accuracies from the \texttt{BERT} baseline. 

At first glance, the baseline results are surprising, since \texttt{BERT} is supposed to have \emph{no specific knowledge of source code}. These concerns are somewhat alleviated by the lack of localization of code properties in \texttt{BERT} layers. 
In addition, tasks such as TYP show a clear advantage for all pre-trained source code models, where they exhibit competitive accuracies with fewer samples, particularly \texttt{GraphCodeBERT}. 


We hypothesize that, rather than \texttt{BERT} possessing competitive knowledge of source code, the most likely cause is that the source code models were not substantially as effective in encoding the probed code characteristics as was expected. Further investigation on more tasks could confirm this; if confirmed this would also suggest that more effective pre-training procedures should be explored.




\section{Conclusion \& Future work}



As more pre-trained code models are introduced to the SE community at a rapid pace, through IDE extensions, plugins, and web engines, e.g. Tabnine, IntelliCode \cite{svyatkovskiy2020intellicode}, TransCoder \cite{lachaux2020unsupervised}, and more recently Github Copilot \cite{DBLP:journals/corr/abs-2107-03374}---it becomes imperative for us to be aware of their capabilities and drawbacks. In order to do so, we probe into four publicly-available pre-trained source code models, surveying a diverse set of code characteristics with relevant and representative {probing tasks}. 

We show how probes help us to gauge the strengths and weaknesses of a model, to understand the role played by the individual hidden layers in model performance, to verify the linear extractability of properties, to get insight on a model's sample efficiency for a given task, and overall to peek into the ``black boxes'' that are large-scale pre-trained models. 

With our initial probes, we were surprised to notice the slim margin of difference in performance between models that should have no knowledge of code and models that do. This clearly needs to be investigated further, and, if confirmed, would suggest that there is room for research in more advanced pre-training techniques for source code models, so that they can effectively leverage their knowledge of source code. 


As future work, we plan to construct further probing tasks evaluating additional source code characteristics, while adding more tasks based on structure, syntax and semantics. Furthermore, we intend to report a benchmark-style comparison study of additional pre-trained models such as \texttt{CuBERT} \cite{DBLP:journals/corr/abs-2001-00059}, \texttt{C-BERT} \cite{DBLP:journals/corr/abs-2006-12641}, \texttt{PLBART} \cite{ahmad2021unified}, and \texttt{Codex} \cite{DBLP:journals/corr/abs-2107-03374}. 

We plan to further extend our suite of probes to more comprehensive source code properties, including context-based probes for applications such as code search and summarization, and extend it to several languages, in order to support a broader array of pre-trained code models. In the long run, a suite of probing tasks could be used to evaluate novel pre-trained source code models, thereby forming a pseudo-benchmark during the development phase, making sure that these models do encode important source code characteristics.

\textit{Reproducibility.} All code, datasets, and experimental results are made available online for replication purposes on github\footnote{https://github.com/giganticode/probes}.

\bibliography{conference}{}
\bibliographystyle{plain}

\end{document}